\documentclass[twocolumn]{revtex4}
\usepackage[dvips]{graphicx}
\usepackage{float}
\begin{document}

\title{Vortex core transitions in superfluid ${}^3$He in globally anisotropic aerogels}

\author{Kazushi Aoyama${}^{1,2}$ and Ryusuke Ikeda${}^1$}

\affiliation{${}^1$ Department of Physics, Kyoto University, Kyoto 606-8502, Japan \\
${}^2$ Young Researcher Development Center, Kyoto University, Kyoto 606-8302, Japan
}

\date{\today}

\begin{abstract}
Core structures of a single vortex in A-like and B-like phases of superfluid ${}^3$He in uniaxially compressed and stretched aerogels are studied by numerically solving Ginzburg-Landau equations derived microscopically. It is found that, although any uniaxial deformation leads to a wider A-like phase with the axial pairing in the pressure-temperature phase diagram, the vortex core states in the two phases in aerogel depend highly on the type of deformation. In a compressed aerogel, the first-order vortex core transition (VCT) previously seen in the bulk B phase appears at {\it any} pressure in the B-like phase while no strange vortex core is expected in the corresponding A-like phase. By contrast, in a stretched aerogel, the VCT in the B-like phase is lost while another VCT is expected to occur between a nonunitary core and a polar one in the A-like phase. Experimental search for these results is hoped to understand correlation between superfluid $^3$He and aerogel structure. 
\end{abstract}

\maketitle

\section{Introduction} 

Superfluid ${}^3$He is in a spin triplet Cooper pairing state with a $p$-wave orbital symmetry \cite{VW}. Recent research interests have been directed toward superfluid ${}^3$He in an aerogel with highly porous structure composed of silica strands to search for possible impurity-induced or anisotropy-induced novel superfluid states \cite{Halperin,IA,Lee,AI06}. At least, two superfluid states, A-like and B-like phases, seem to appear in aerogel and their pairing states are the same as those of the bulk A and B phases, i.e., the axial \cite{AI06,AI07,Volovik} and isotropic pairings, respectively. On the other hand, by introducing some anisotropy over large scales in aerogel, a change in pairing state might occur. 

In aerogel, {\it locally} anisotropic scattering events due to silica strands may stabilize the A-like phase, although this effect is mostly cancelled by an impurity-induced reduction in the strong-coupling (SC) correction \cite{AI07,A1A2,Moon} necessary for stabilizing the axial pairing. Recent studies have demonstrated that a {\it global} anisotropy existing over large scales helps to stabilize the A-like phase because it promotes an orientational ordering of the orbital axis ${\bf l}$ of the axial pairing state and may screen quenched disorder effects of aerogel on ${\bf l}$. For instance, in a uniaxially compressed aerogel, the ${\bf l}$ vector of the A-like phase tends to be aligned along the compressed direction while, in a uniaxially stretched aerogel, ${\bf l}$ should lie in the plane perpendicular to the stretched direction. Such a control of the ${\bf l}$ texture due to a deformation seems to have been verified through recent NMR measurements \cite{Bunkov,Kunimatsu}. In addition, as the stretched deformation is enhanced, the novel {\it polar} pairing state may be realized in a narrow temperature region just below the superfluid transition temperature $T_c(P)$ \cite{AI06}. Although the polar pairing state has not been detected yet experimentally, a wider A-like phase and an emergence of the polar pairing state might be indicated in spatially local regions as a consequence of the multiplicity of components of the order parameter in $^3$He. 
                                                                               
In this work, core states of an isolated vortex in superfluid $^3$He in globally anisotropic aerogels are examined based on a Ginzburg-Landau (GL) free-energy functional ${\cal H}_{\rm GL}$ derived microscopically. By assuming a boundary condition consistent with the radially symmetric coreless vortex \cite{MH}, we find that a polar-core vortex and the resulting vortex core transition (VCT) in the A-like phase are obtained in a uniaxially stretched aerogel. In contrast, in the A-like phase in uniaxially compressed aerogel, the ${\bf l}$ vector tends to align along the anisotropy axis so that no intriguing vortex core structure results in. Therefore, regarding core structures in the A-like phase, we will limit ourselves to possible situations in the A-like phase in uniaxially {\it stretched} aerogel. We also examine possible vortex core structures in the B-like phase and find that the VCT found in the bulk B phase \cite{Thuneberg,SV}, which occurs between a nonaxisymmetric structure and an axisymmetric one with A-phase core, survives in a uniaxially compressed aerogel even in the low-pressure limit while such a VCT accompanied by the A-phase core is not realized in the stretched aerogel. 

\section{Ginzburg-Landau Theory}

Our starting point is the weak-coupling BCS Hamiltonian with a $p$-wave pairing interaction term and a term describing impurity scatterings brought by the aerogel structure \cite{IA,AI06}. The latter term is expressed in the simple nonmagnetic and potential scattering form 
\begin{equation}
{\hat H}_{\rm imp} = \int d^3r \, u({\bf r}) \, {\hat n}({\bf r}), 
\end{equation}
where ${\hat n}$ is the quasiparticle number density. The scattering potential $u$ has zero mean and yields the ensemble 
\begin{equation} \label{Eq:aisoimp}
{\overline {|u_{\bf k}|^2}} = u_0^2 (1+\delta_u(\hat{\bf k}\cdot\hat{z})^2) 
\end{equation}
at a {\it fixed} ${\bf k}$, where $u_{\bf k}$ is the Fourier transform of $u({\bf r})$, ${\hat {\bf k}}={\bf k}/k_F$, the overbar denotes the random average, and a small parameter $\delta_u$ measuring the nature of a global anisotropy is negative (positive) for uniaxially stretched (compressed) aerogels \cite{AI06}. Within the simplest Born approximation, the relaxation rate becomes $\tau^{-1} = 2 \pi N(0) u_0^2 (1 + \delta_u/3)$, where $N(0)$ is the density of state per spin on the Fermi surface. Although, as a model for real aerogels, $u_{\bf k}$ should have more complicacted ${\bf k}$ dependence, other ${\bf k}$ dependence expressing the {\it local} anisotropy of aerogel will be neglected hereafter by assuming the global anisotropy to screen the quenched disorder effects of aerogel on ${\bf l}$. The GL free-energy functional ${\cal H}_{\rm GL}$ has been derived elsewhere \cite{AI06} from the BCS Hamiltonian with the anisotropic impurity scattering potential defined by Eq. (\ref{Eq:aisoimp}) and is expressed as 
\begin{eqnarray}\label{Eq:GL1}
{\cal H}_{\rm GL} &=& \int d^3r \biggl[ \Bigl(\alpha_0 \delta_{i,j} + \alpha_z \delta_{i,z} \delta_{j, z} \Bigr) A_{\eta,i}^\ast A_{\eta,j} \nonumber\\
&+& K A_{\eta,i}^\ast \Bigl(-\partial_i \partial_j  - \frac{1}{2} \nabla^2 \delta_{i,j} \Bigr)   A_{\eta,j} + \beta_1 |A_{\eta,i}A_{\eta,i}|^2 \nonumber \\
&+& \beta_2 (A_{\eta,i}^{\ast}A_{\eta,i})^2 
+ \beta_3 A_{\eta,i}^{\ast}A_{\nu,i}^{\ast}A_{\eta,j}A_{\nu,j} \nonumber \\ 
&+& \beta_4 A_{\eta,i}^{\ast}A_{\nu,i}A_{\nu,j}^{\ast}A_{\eta,j} + \beta_5 A_{\eta,i}^{\ast}A_{\nu,i}A_{\nu,j}A_{\eta,j}^{\ast} \biggr],
\end{eqnarray}
where $A_{\eta, i}$ is the order-parameter field with a spin index $\eta$ and an orbital one $i$ which denote the Cartesian coordinate axes, 
\[
\alpha_0=\frac{N(0)}{3}\bigg[\ln\Big(\frac{T}{T_{c0}}\Big)+\psi\Big(\frac{1}{2} + d_T \Big)-\psi\Big(\frac{1}{2}\Big) \bigg] + \frac{3 \, \alpha_z}{16},
\]
\[
\alpha_z=\frac{16 N(0) \, d_T \, \delta_u}{45} \psi^{(1)}\Big(\frac{1}{2}+d_T \Big), 
\]
\[
K=-\frac{N(0)}{15}\Big(\frac{v_{\rm F}}{4\pi T}\Big)^2 \, \psi^{(2)}\Big(\frac{1}{2}+ d_T\Big),
\]
\[
 -2\beta_1 = \beta_3 = \beta_2+\varepsilon_{\rm imp}=\beta_4+\varepsilon_{\rm imp}=-\beta_5+\varepsilon_{\rm imp} \equiv \beta_0^{(d)}(T),
\]
\[
\beta_0^{(d)}(T)=-\frac{\beta_0(T)}{7\zeta(3)}\psi^{(2)}\Big(\frac{1}{2}+d_T\Big),
\]
\begin{equation}
\varepsilon_{\rm imp}=\frac{5 \, d_T}{18} \frac{\beta_0(T)}{7\zeta(3)} \psi^{(3)}\Big(\frac{1}{2}+d_T\Big),
\end{equation}
where $T_{c0}$ is the superfluid transition temperature in bulk, $d_T=1/(4 \pi T \tau)$, $\beta_0(T)=7\zeta(3)N(0)/(240 \pi^2T^2)$, $v_{\rm F}$ is the Fermi velocity, $\psi^{(n)}(z)$ is the $n$th order digamma function, and $\psi(z)=\psi^{(0)}(z)$. 
In Eq. (\ref{Eq:GL1}), any term describing quenched disorder effect on $A_{\eta,i}$ of the {\it locally} anisotropic aerogel has been dropped by favoring effects of a global anisotropy of aerogel. Further, higher order contributions in $d_T \, \delta_u$ are also dropped in ${\cal H}_{\rm GL}$ by assuming them to be negligibly small. 

The SC correction $\delta\beta_j$ to $\beta_j$ needs to be incorporated. Effects of an aerogel on $\delta\beta_j$ have been studied thoroughly in Ref.\cite{AI07} by examining a general form of $\delta\beta_j$, and it is found that the impurity effect other than the relaxation rate $\tau^{-1}$, which is a contribution to the quasiparticle effective interaction \cite{AI07}, leads to a reduction in $|\delta\beta_j|$. Hereafter, the expression of $\delta\beta_j$ in the spin-fluctuation model \cite{BSA} will be used for simplicity. In this model, we have $\delta\beta_3= (\delta\beta_2+5\delta\beta_1)/6$, $\delta\beta_4=\delta\beta_3+5\delta\beta_1$, and $\delta\beta_5= 7\delta\beta_1$, and 
\[
\delta\beta_1=-3.3\times10^{-3} \, \delta \, \beta_0(T) \frac{T}{T_{c0}}\sum_m[(D^{(d)}_1(m)]^2, 
\]
\begin{eqnarray}
\delta\beta_2&=&\delta\beta_1\sum_m\{9[D^{(d)}_2(m)]^2-6D^{(d)}_1(m)D^{(d)}_2(m)\nonumber\\
&&-2[D^{(d)}_1(m)]^2\}/\big(\sum_m[D^{(d)}_1(m)]^2\big),\nonumber
\end{eqnarray}
\begin{eqnarray}
D^{(d)}_1(m)&=&\frac{1}{2}\Big[\frac{1}{|m|}+\frac{1}{|m|+2 d_T}\Big]\nonumber\\
&&\times \Big[\psi\Big(\frac{1}{2}+|m|+d_T\Big)-\psi\Big(\frac{1}{2}+d_T\Big)\Big], \nonumber
\end{eqnarray}
\begin{equation}
D^{(d)}_2(m)=\frac{1}{2}\psi^{(1)}\Big(\frac{1}{2}+|m|+d_T\Big), 
\end{equation}
where $\delta$ is defined in Eq. (3.10) of Ref. \cite{BSA} and is scaled with $T_{c0}/E_F$ and $E_F$ is the Fermi energy. Then, we express $\delta = \eta \, T_{c0}/E_F$ and take $\eta$ as a parameter measuring the strength of the SC correction. Below, the value 300 will be used as $\eta$ of the bulk liquid together with $T_{c0}(P)$ (Ref. \cite{VW}) and $E_F(P)$ (Ref. \cite{Kuroda}) data. In the aerogel case, a smaller $\eta$ value will be assumed to mimic the reduction in the SC correction in aerogel \cite{AI07}.

Below, we study the structure of an isolated vortex parallel to the $z$ axis by solving the GL differential equations under a suitable boundary condition. Following Salomaa and Volovik's approach \cite{SV}, we solve the equations in the case with an axisymmetric vortex and then, determine possible vortex structures by introducing a nonaxisymmetric part of each $A_{\eta,i}$ as a perturbation and examining the stability of the axisymmetric solution against nonaxisymmetric ones. 
To proceed, it is convenient to rewrite ${\cal H}_{\rm GL}$ by representing $A_{\eta,i}({\bf r})$ in terms of the cylindrical frame as follows: 
\begin{equation}\label{Eq:relation}
A_{\eta,i}({\bf r})=\Delta(T) \sum_Q \, \sum_{\mu,\nu} \, \lambda^\mu_\eta\lambda^\nu_i \, C^{(Q)}_{\mu,\nu}(r) \, e^{i Q \phi}\,  e^{i(m-\mu-\nu)\phi},
\end{equation}                                                                 
where $\Delta(T)$ is the energy-gap amplitude far from the vortex axis, $m$ is the circulation quantum number, $Q$ is an integer, $\mu$ and $\nu$ specify the projections of the spin and the orbital angular momentum of the Cooper pair, respectively, and take three values (+1, 0, -1) with $\lambda^{\pm}_i=(\hat{x}_i\pm i\hat{y}_i)/\sqrt{2}$, $\lambda^0_i=\hat{z}_i$. Note that, hereafter, all components of $C^{(Q)}_{\mu,\nu}(r)$ have only radial dependences and that axisymmetric and nonaxisymmetric parts of $A_{\eta,i}$ correspond to $C^{(Q)}_{\mu,\nu}(r)$ with $Q=0$ and with nonzero $Q$, respectively. Equation (\ref{Eq:relation}) shows that the axisymmetric component $C^{(0)}_{\mu,\nu}$ with $m-\mu-\nu \neq 0$ does not become nonzero at the vortex center where $r=0$ since $\phi$ becomes multi-valued there while the components with $m-\mu-\nu = 0$ can exist at $r=0$. In contrast to the case of $s$-wave superfluid where a vortex always has a normal core, an axisymmetric vortex in superfluid ${}^3$He can have nonvanishing superfluid components with $m-\mu-\nu = 0$ at the vortex center instead of the normal core so as to gain the condensation energy. Hereafter, a vortex whose center is occupied with a superfluid component X differing from the components realized far from the vortex center will be called as "a X-core vortex." 

In rewriting ${\cal H}_{\rm GL}$ in terms of Eq. (\ref{Eq:relation}), nonaxisymmetric ($Q\neq 0$) components $C^{(Q)}_{\mu,\nu}$ with the same $|Q|$ couple with each other, while those with different $|Q|$ are not mixed at the lowest order in the correction $C^{(Q\neq 0 )}_{\mu,\nu}$. Thus, we rewrite the GL free-energy functional per unit length in the $z$-direction ${\cal H}_{\rm GL}^{\rm (2D)}$ as follows:   
\[
{\cal H}_{\rm GL}^{\rm (2D)}= 2\pi K |\Delta(T)|^2  \int_0^{\infty} \tilde{r} \, d\tilde{r} \, f(\tilde{r})
\]
\begin{eqnarray}\label{Eq:GL2}
 f(\tilde{r})&=&\sum_{q= \, \pm Q,0 }\bigg[(\tilde{\alpha}_0+\tilde{\alpha}_z \delta_{\nu,0}) C^{(q)}_{\mu,\nu}C^{(q)*}_{\mu,\nu} \nonumber\\
&+&|\frac{\partial}{\partial \tilde{r}}C^{(q)}_{\mu,\nu}|^2 +\bigg(\frac{m-\mu-\nu+q}{\tilde{r}}\bigg)^2|C^{(q)}_{\mu,\nu}|^2 \nonumber\\
&+&\Big(|\nu|\frac{\partial}{\partial \tilde{r}}C^{(q)}_{\mu,\nu}-\nu\frac{(m-\mu-\nu+q)}{\tilde{r}}C^{(q)}_{\mu,\nu}\Big) \nonumber\\
&&\times \Big(|\nu'|\frac{\partial}{\partial \tilde{r}}C^{(q)*}_{\mu,\nu'}-\nu'\frac{(m-\mu-\nu'+q)}{\tilde{r}}C^{(q)*}_{\mu,\nu'}\Big) \bigg] \nonumber\\
&+&\sum_{q_i= \, \pm Q,0} [ \, \tilde{\beta}_1 C^{(q_1)*}_{\mu,\nu}C^{(q_2)*}_{-\mu,-\nu}C^{(q_3)}_{\rho,\kappa}C^{(q_4)}_{-\rho,-\kappa} \nonumber\\
&+&\tilde{\beta}_2 C^{(q_3)}_{\mu,\nu}C^{(q_1)*}_{\mu,\nu}C^{(q_4)}_{\rho,\kappa}C^{(q_2)*}_{\rho,\kappa} + \tilde{\beta}_3 C^{(q_3)}_{\mu,\nu}C^{(q_1)*}_{\mu,\kappa}C^{(q_4)}_{\rho,-\nu}C^{(q_2)*}_{\rho,-\kappa} \nonumber \\ 
&+& \tilde{\beta}_4 C^{(q_1)*}_{\mu,\kappa}C^{(q_3)}_{\nu,\kappa}C^{(q_2)*}_{\nu,\rho}C^{(q_4)}_{\mu,\rho} + \tilde{\beta}_5 C^{(q_3)}_{\mu,\nu}C^{(q_4)}_{-\mu,\kappa}C^{(q_1)*}_{\rho,\nu}C^{(q_2)*}_{-\rho,\kappa} ], \nonumber\\
\end{eqnarray}
where summations for $q_i$ are carried out under the constraint $q_1+q_2-q_3-q_4=0$, $\tilde{\alpha}_n \equiv \alpha_n/|\alpha_0+\alpha_z O_z^2|$ $(n=0,z)$, $\tilde{\beta}_i \equiv  |\Delta(T)|^2 \beta_i/|\alpha_0+\alpha_z O_z^2|$, ${\tilde r}\equiv r/\xi_{\rm GL}(T)$, $\xi_{\rm GL}(T) \equiv (K/|\alpha_0 + O_z^2 \alpha_z|)^{1/2}$, and $|O_z|$ is the magnitude of $A_{\eta,z}$ far from the vortex center (see below).

We set the boundary condition at ${\tilde r}_b$ far from the vortex center and numerically solve the differential equations obtained by varying Eq.(\ref{Eq:GL2}) with respect to $C^{(Q)}_{\mu,\nu}$. As indicated in Eq. (\ref{Eq:GL2}), we can examine the instability of an axisymmetric ($Q=0$) solution against nonaxisymmetric corrections $C^{(Q\neq 0)}_{\mu,\nu}$ at each $|Q|$ value independently. Furthermore, any component with $|Q|\geq 3+m$ cannot survive at the vortex center since $\phi$ becomes multi-valued ($\pm |Q|+m-\mu-\nu \neq 0$) there. Since we are interested in the order-parameter structure close to the vortex center, for brevity, we will incorporate only corrections with $|Q| \leq 2+m$ so that higher $|Q|$ components are neglected.  
Below, we show results of our calculation for $m=1$ vortices in deformed B-like phases and the uniaxially stretched A-like phase by assuming each of them to be a $v$ vortex in which all components of $C^{(Q)}_{\mu,\nu}$ are real. In obtaining the results, the parameters $1/(2\pi \tau)=0.135$ (mK) and $\eta=266$ are used in the aerogel case, and the vortex radius ${\tilde r}_b=20$ is assumed.

\section{Vortex core transition in B-like phases in anisotropic aerogel}

First, a possible VCT in the B-like phase in uniaxially deformed aerogels will be considered. The uniaxial deformation also affects $A_{\eta,i}$ far from the vortex core, and thus, the boundary condition we choose on $A_{\eta,i}$ far from the vortex core is 

\begin{equation}
A_{\eta ,i}(|{\tilde {\bf r}}|=\infty)=\Delta_B(T) \, e^{i \phi}\big[O_{xy}\delta_{\eta,i} (1 - \delta_{i,z}) + O_{z} \delta_{\eta,z}\delta_{i,z} \big],
\end{equation}
or equivalently, $C^{(0)}_{-+}=C^{(0)}_{+-}=O_{xy}, \, C^{(0)}_{00}=O_z$. By keeping the normalization $2 O_{xy}^2 + O_z^2=1$, $O_z$ is determined by minimizing the free energy in the vortex-free situation. 

\begin{figure}[t]
\begin{center}
\includegraphics[scale=0.55]{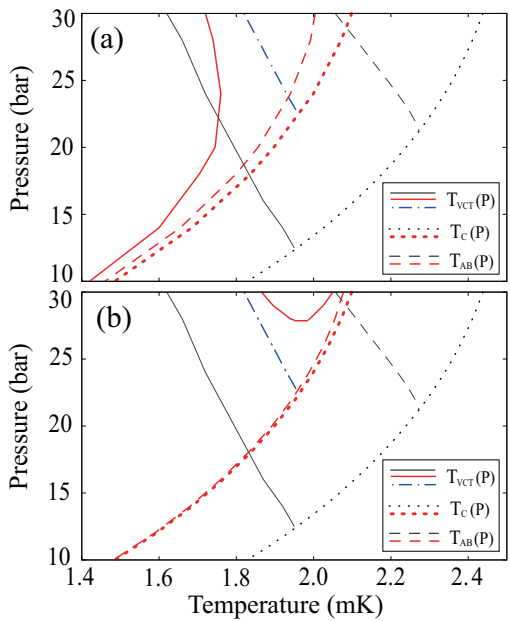}
\caption{(Color online) Calculated pressure to temperature phase diagram consisting of the corresponding VCT curve $T_{\rm VCT}(P)$ (solid curve) in the bulk B phase or B-like phase, the superfluid transition curve $T_c(P)$ (dotted one), and the corresponding A-B transition curve $T_{\rm AB}(P)$ (dashed one) of liquid $^3$He in bulk [thin (black) curves] and an uniaxially deformed aerogel [thick (red) curves]. For comparison, we also show the VCT curve [dashed-dotted (blue) curve] in a globally {\it isotropic} aerogel where the resulting $T_c(P)$ cannot be distinguished from that [thick (red) dotted curve] in the anisotropic ones and thus, is not drawn in the figures. The parameters $\delta_u=0.01$ and $\delta_u=-0.003$ are used for (a) compressed and (b) stretched aerogels, respectively, and $(2 \pi \tau)^{-1} = 0.135$ (mK) was assumed in both of (a) and (b). The A-phase-core vortex is stable at higher temperatures than the VCT curve. The VCT curve in the isotropic aerogel shifts to higher temperatures than that in bulk, reflecting the reduction in the SC correction. As shown in (a), the A-phase-core vortex is stabilized by the small compression and survives even in the low-pressure limit, while in (b), it tends to be lost due only to a small stretch. \label{Fig:bw-vt}}
\end{center}                                                                                                                                                                                                                                                                                                                                                                                                                                                                                                                                              
\end{figure}

In the bulk B phase where $1/\tau =0$ and $O_{xy}=O_z=1/\sqrt{3}$, we obtain two solutions of a single vortex, an axisymmetric vortex and a nonaxisymmetric one with $|Q|=2$ components, in agreement with the previous results \cite{Thuneberg,SV}. The former vortex gains a large $|C^{(0)}_{0 \, +}|$ and a small $|C^{(0)}_{+ \, 0}|$ at the vortex center which correspond to the A-phase component with ${\bf l} \parallel z$ and the $\beta$-phase \cite{VW} one, respectively. Thus, this structure is called an A-phase-core vortex. On the other hand, the latter vortex has nonvanishing four components $C^{(0)}_{0 \, +}$, $C^{(-2)}_{0 \, -}$, and $C^{(0)}_{+ \, 0}\simeq C^{(-2)}_{- \, 0}$ at the vortex center, forming a twofold symmetric vortex core state, and thus, is called a double-core vortex. We also compare the total free energy $\int d\tilde{r}\, \tilde{r} \, f(\tilde{r})$ of the axisymmetric vortex with that of the nonaxisymmetric vortex and find that the axisymmetric A-phase-core vortex is stable at higher temperatures. The resulting transition (VCT) curve between the two vortex core states is denoted by a thin (black) solid curve in Fig. 1. 
The stability of the A-phase-core vortex against the double-core vortex results from the large gain in the condensation energy of its core phase overcoming the energy cost due to the spacial variation in the order parameter larger than that for the double-core vortex \cite{Thuneberg}. The solid black curve in Fig. 1 remains nearly straight, reflecting the bulk A phase stabilized by the SC correction, namely, the pressure dependence of $T_{c0}/E_F$ in $\delta$ in the spin-fluctuation model. However, the experimentally obtained VCT curve has shown an upturn near $T_c$ (Ref. \cite{VTexp}) in contrast to the present result. As a possible origin of this discrepancy, let us discuss our truncation of the summation for $Q$ in Eq. (\ref{Eq:relation}) which is carried out in obtaining Eq. (\ref{Eq:GL2}) (see the final paragraph in Sec. II). In the $|Q|=2$ nonaxisymmetric vortex where the twofold symmetry is retained, we may need to include components with larger even $|Q|$ values \cite{SV}.  
Since such higher $|Q|$ components do not appear at ${\tilde r}=0$ and hardly affect the order parameter at the vortex core, we expect that the energy gain of the $|Q|=2$ double-core vortex due to the higher $Q$ components mainly stems from the spacial variation in the order parameter far from the core which is relatively insensitive to temperature. Thus, inclusion of the higher $Q$ components would merely shift the VCT curve to a higher pressure while keeping the shape of the VCT curve unchanged and thus, would not lead to resolving the discrepancy on the VCT curve near $T_c(P)$. Since the problem concerning the discrepancy is not easy to resolve and needs further studies, we will not discuss the problem here. Below, we will show how the stability region of the A-phase-core vortex in the phase diagram is altered by aerogels. 

In a globally isotropic aerogel ($\delta_u=0$) where the region of the A-like phase is invisible in the parameter range used for our numerical analysis, the corresponding VCT curve denoted by a dashed-dotted (blue) one in Fig. 1 shifts to higher temperatures than the bulk VCT curve, as a result of the reduction in the SC correction. In globally anisotropic aerogel, the shape of the VCT curve in the B-like phase is drastically altered by only a small uniaxial deformation, depending on whether it is a compression or a stretch [see Figs. 1(a) and 1(b)]. A small uniaxial compression, measured by $\delta_u=0.01$, expands not only the A-like phase region but also the region of A-phase-core vortex while quite a small uniaxial stretch, measured by $\delta_u=-0.003$, suppresses the region of the A-phase-core vortex although the A-like phase itself expands due to the stretch. Such dependence of the A-phase-core vortex on the type of deformation of aerogel is qualitatively reasonable because, just as in the vortex-free A-like phase induced by a compression, ${\bf l}$ is aligned along the vortex axis in the A-phase core of the bulk B phase \cite{Thuneberg,SV}. The A-phase-core vortex region in the stretched case becomes narrower in temperature, since the effect of suppression of the A-phase-core state due to the uniaxial stretch is so strong to overwhelm the SC correction just below the A-B transition curve $T_{\rm AB}(P)$. With increasing $|\delta_u|$, the VCT curve is extended down to lower temperatures in the compressed case \cite{AI09} while, in the stretched case, the region of the A-phase-core vortex is suppressed and vanishes for larger $|\delta_u|$ than the value $\delta_u=-0.006$.  
                                                                                                                                                                                                                                                                                                                                                                                                                                                                                                                                                                                                                                                                                                                                                                                                                       
\section{Vortex core transition in A-like phase in stretched aerogel}

\begin{figure}[t]
\includegraphics[scale=0.46]{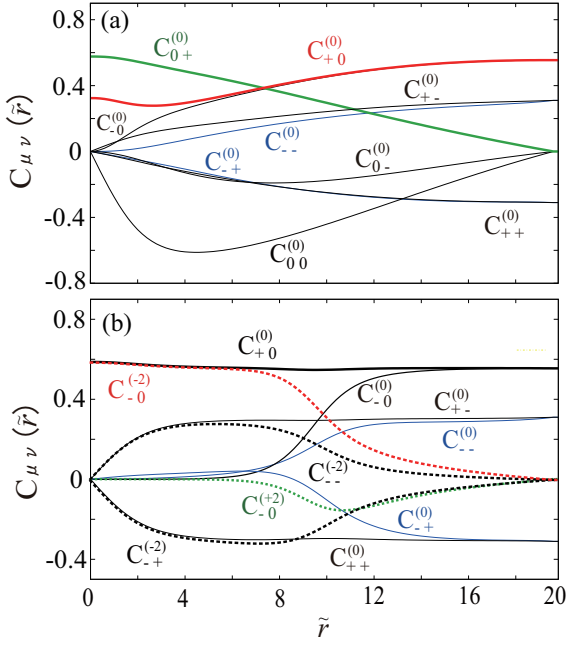}
\caption{(Color online) Radial dependences of $C^{(Q)}_{\mu,\nu}({\tilde r})$ of (a) an axisymmetric vortex and (b) a nonaxisymmetric one with $|Q|=2$ components at T=2.040(mK) and P=28 (bar) in the A-like phase in the uniaxially stretched aerogel with $\delta_u=-0.03$. In (a), $C_{+ 0}^{(0)}$ and $C_{0 +}^{(0)}$ remain finite at ${\tilde r}=0$ which correspond to the $\beta$-phase component and that of MHV core, respectively. In (b), the components $C^{(Q)}_{0,\nu}({\tilde r})$ are identically zero and other negligibly small components are not shown. Further, the dashed curves imply $Q \neq 0$ components. At ${\tilde r}=0$, just the two components $C^{(0)}_{+0}$ and $C^{(-2)}_{-0}$ remain finite, implying a polar core $A_{\eta,i}(r=0) = a_{x,z} \delta_{\eta,x} \delta_{i,z}$. \label{Fig:op}}
\end{figure}                                                                   

Next, we investigate vortex structures in A-like phases in aerogels deformed along the vortex axis by paying our attention to the deformation-induced instability of a coreless Mermin-Ho vortex (MHV) (Ref. \cite{MH}) in the bulk A phase. In uniaxially compressed aerogels, the ${\bf l}$ vector tends to become parallel to the vortex axis and the boundary condition of MHV far from the vortex center is incompatible with the orientation of ${\bf l}$. Once the boundary condition compatible with ${\bf l} \parallel z$ is chosen, the uniaxial compression does not affect vortex structures any longer and, instead, the pure phase vortex will appear. We focus hereafter on the uniaxially stretched case in which ${\bf l}$ is perpendicular to the vortex axis far from the vortex center. 
As the boundary condition, we take the following condition compatible with that of the dipole-locked MHV                                                                                                                                                                                                                                                                                                                                                                                                                                                                                                                                                                     
\begin{eqnarray}\label{Eq:alike}
A_{\eta ,i}(|{\bf {\tilde r}}|=\infty)&=&\Delta_A(T) \, e^{i \phi} {\bf d}_\eta  ({\bf m} \, O_z  -i \, {\bf n} \, O_{xy})_i \nonumber\\
{\bf d}&=&\hat{x} \, \cos\phi  + \hat{y} \, \sin\phi ,\nonumber\\
{\bf m}&=&\hat{z}, \, {\bf n}=\hat{\phi} 
\end{eqnarray}
with the normalization $O_z^2 + O_{xy}^2=1$. In the representation in Eq. (\ref{Eq:relation}), we have $C^{(0)}_{--}=C^{(0)}_{+-}=-C^{(0)}_{++}=-C^{(0)}_{-+}=O_{xy}/2$ with $C^{(0)}_{+0}=C^{(0)}_{-0}=O_z/\sqrt{2}$ or, equivalently, the radial ${\bf l}$ {\it and} ${\bf d}$ vectors. In the A-like phase in a uniaxially stretched aerogel, we obtain three solutions, an axisymmetric vortex, a nonaxisymmetric vortex with $|Q|=1$ components, and a nonaxisymmetric vortex with $|Q|=2$ components. The $|Q|=1$ nonaxisymmetric vortex always has a higher value of the free energy $\int_0^{\infty}d\tilde{r} \, \tilde{r}f(\tilde{r})$ than two other solutions, and thus, it will not be discussed hereafter. Below, we will show results of our calculation in the case with $\delta_u=-0.03$. The resulting phase diagram is shown in Fig. \ref{Fig:vt-abm}, where a dashed curve denotes $T_{\rm AB}(P)$ and the polar pairing state occurs in the narrow region sandwiched between $T_c(P)$ (dotted) curve and a dashed-dotted one. A solid (red) curve and the inset in Fig. \ref{Fig:vt-abm} will be explained later.                                                                                                                    
\begin{figure}[t]
\includegraphics[scale=0.46]{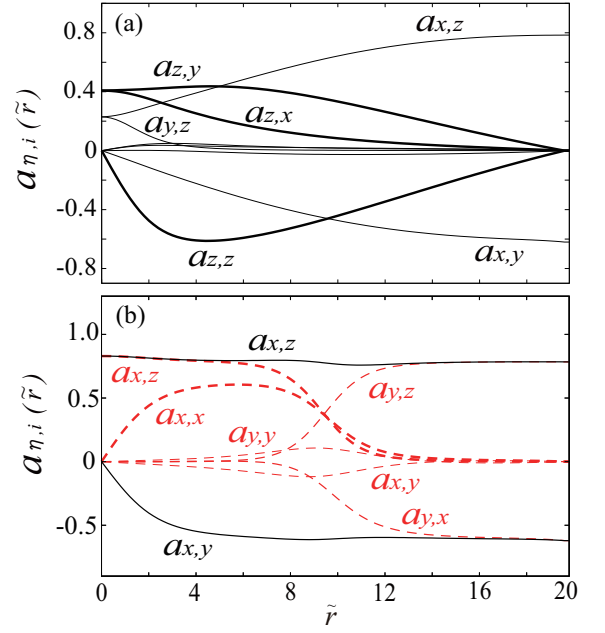}
\caption{(Color online) Spacial variations in the order parameter $A_{\eta ,i}$ of (a) the axisymmetric vortex and (b) the nonaxisymmetric one on the $x$ axis (black curves) and $y$ axis (red dashed ones). In (a), the vortex center is occupied with the $\beta$ phase $a_{x,z} \, ({\hat x}+ i \, {\hat y})_\eta \, \delta_{\i,z}$ and the component of the dipole-locked MHV center $a_{z,y}\, \delta_{\eta,z} \, ({\hat x}+ i \, {\hat y})_i$. In (b), $a_{\eta,i}$'s start varying around ${\tilde r}=10$ on the $y$ axis, corresponding to the ${\bf d}$ texture (see the text below), and a polar state $a_{x,z} \, \delta_{\eta,x} \delta_{i,z}$ is realized at the vortex center. The three thin solid curves in (a) with extremely small magnitudes denote $a_{y,y}$ (top), $a_{y,x}$ (middle), and $a_{x,x}$ (bottom). Negligibly small contributions on the $x$ axis are not shown in (b). \label{Fig:op-xy}}
\end{figure}  

Figure \ref{Fig:op} shows the radial dependences of $C^{(Q)}_{\mu,\nu}({\tilde r})$ for (a) an axisymmetric vortex and (b) a nonaxisymmetric one with $|Q|=2$ components at $T=2.040$ (mK) and $P=28$ (bar), and Figs. \ref{Fig:op-xy}(a) and \ref{Fig:op-xy}(b) show the corresponding spacial variations in the order-parameter components in $A_{\eta ,i}(r,\phi)$ on the $x$ axis [solid (black) curve] and the $y$ axis [dashed (red) one], where
\begin{eqnarray}\label{Eq:alike_xy}
A_{\eta ,i}(r, \phi)&=&\Delta_{\rm A}\bigl[\delta_{\eta,x}(a_{x,x} \, e^{i \phi} \, \hat{x}+i \, e^{-i \phi} \, a_{x,y} \, \hat{y}+a_{x,z} \, \hat{z})_i \nonumber\\
&&+\delta_{\eta,y}(i \, a_{y,x} \, e^{-i \phi} \, \hat{x}+a_{y,y} \, e^{i \phi} \, \hat{y}+i \, a_{y,z} \, \hat{z})_i  \nonumber\\
&&+\delta_{\eta,z}( a_{z,x} \, \hat{x}+i \, a_{z,y} \, \hat{y}+e^{i \phi} \, a_{z,z} \, \hat{z})_i \bigr]
\end{eqnarray}
on the $x$ ($\phi=0$) and $y$ ($\phi=\pi/2$) axes.
The axisymmetric vortex shown in Fig. \ref{Fig:op}(a) has nonvanishing $C_{+ 0}^{(0)}$ and $C_{0 +}^{(0)}$ components at the vortex center ${\tilde r}=0$ which correspond to the nonunitary $\beta$-phase component $ a_{x,z} ({\hat x} +i \, {\hat y} )_\eta \, \delta_{z,i}$ and the A-phase one $a_{z,x} \, \delta_{\eta,z}( \hat{x}+i \, \hat{y})_i$ in Fig. \ref{Fig:op-xy}(a), respectively. As one can see in Fig. \ref{Fig:op-xy}(a), with increasing ${\tilde r}$, $A_{\eta ,i}$ changes from $\delta_{\eta,z}[(a_{z,x} \, \hat{x}+a_{z,z} \, \hat{z})+i \, a_{z,y} \, \hat{y}\,]_i$ to $\delta_{\eta,x}(a_{x,z} \, \hat{z}+i \, a_{x,y} \, \hat{y}\,)_i$, indicating the continuous rotation of the ${\bf l}$ vector and the ${\bf d}$ vector from ${\hat z}$ to ${\hat x}$. This rotational behavior in ${\bf l}$ and ${\bf d}$ is compatible with the texture in the dipole-locked MHV so that we call the axisymmetric vortex the nonunitary Mermin-Ho vortex. As one can infer from the sign of $\alpha_z$ in Eqs. (\ref{Eq:GL1}) and (\ref{Eq:GL2}), the uniaxial stretch ($\delta_u <0$) enhances $a_{\eta, z}$ corresponding to $C^{(0)}_{+0}$. As temperature is lowered so that the uniaxial stretch is less effective, $C^{(0)}_{+0}({\tilde r}=0) $ decreases, while $C^{(0)}_{0+}({\tilde r}=0) $ grows up so that $A_{\eta,i}$ approaches that of MHV.  
It is noted that this axisymmetric vortex has a nonvanishing magnetic moment parallel to the vortex axis arising from the nonunitary $\beta$-phase component $C^{(0)}_{+0} ({\tilde r}=0) $ which ramains finite even at the lowest temperature in the A-like phase, $T_{\rm AB}(P)$. Figure \ref{Fig:op} (b) shows the structure of the nonaxisymmetric vortex, 
where a $|Q|$=2 nonaxisymmetric part, $C^{(-2)}_{-0}$, denoted by a dashed curve remains nonvanishing as well as $C^{(0)}_{+0}$ at ${\tilde r}=0$, and the resulting core state becomes the nonaxisymmetric polar pairing one, $A_{\eta,i} = a_{x,z} \, \delta_{\eta,x} \delta_{i,z}$, as shown in Fig. \ref{Fig:op-xy} (b). 

\begin{figure}[t]
\includegraphics[scale=0.45]{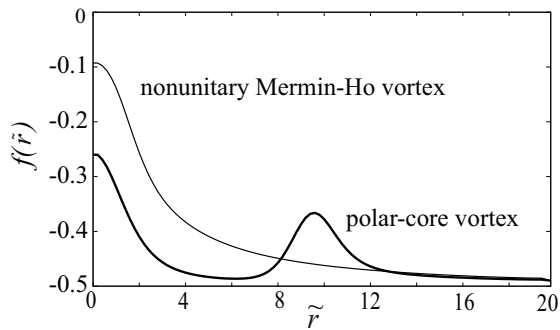}
\caption{(Color online) Comparison of radial dependence of the free energy density between the axisymmetric nonunitary Mermin-Ho vortex and the nonaxisymmetric polar-core vortex. For both the two curves, the free energy density was normalized by $2\pi K|\Delta|^2$ at T=2.040(mK) and P=28 (bar). \label{Fig:edens}}
\end{figure}        

The radial dependence of the free-energy density $f({\tilde r})$ of the polar-core vortex is also shown in Fig. \ref{Fig:edens}, where $f({\tilde r})$ of the nonunitary MHV is shown for comparison. The lower free-energy density of the polar-core vortex close to the vortex center competes with its bump appearing around ${\tilde r}=10$ in $f({\tilde r})$. Comparing the total free energies $\int d\tilde{r}\, \tilde{r} \, f(\tilde{r})$ of the two vortices with each other, we find that the polar-core vortex is more stable than the nonunitary MHV at any temperature in the A-like phase. The stability of the polar-core vortex against the nonunitary MHV, however, depends on the vortex radius $\tilde{r}_b$. When ${\tilde r}_b=50$ is assumed, the axisymmetric vortex has lower free energy than that of the nonaxisymmetric vortex in contrast the case with $\tilde{r}_b=20$. For the intermediate value $\tilde{r}_b=30$, temperature dependences of the free energies of the two vortices at $P=28$ (bar) are shown in the inset of Fig. \ref{Fig:vt-abm}, where the VCT close to 2.02(mK) is realized. The VCT curve obtained for a vortex with the radius $\tilde{r}_b=30$ is denoted by a solid (red) curve in Fig. \ref{Fig:vt-abm} where the polar-core vortex is stable at higher temperatures than the solid curve. Even in a more realistic situation with a fixed $r \equiv R$, where $R$ denotes the averaged spacing between the neighboring vortices, the VCT should be realized because, at a fixed pressure, ${\tilde R}=R/\xi_{\rm GL}(T)$ becomes longer on cooling due to the temperature dependence of $\xi_{\rm GL}(T)$. In this way, it is concluded that a discontinuous VCT from the nonaxisymmetric polar core to the axisymmertric nonunitary core should occur in the A-like phase. 

\begin{figure}[t]
\includegraphics[scale=0.51]{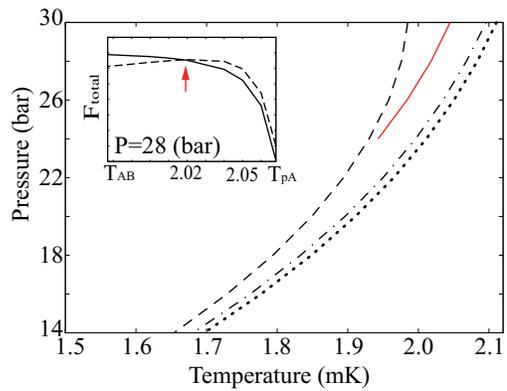}
\caption{(Color online) Calculated pressure to temperature phase diagram consisting of VCT (solid) curve in the A-like phase, $T_c(P)$ (dotted) curve, $T_{\rm AB}(P)$ (dashed) curve, and the polar-A transition (Ref. \cite{AI06}) (dashed-dotted) curve in superfluid $^3$He in the uniaxially stretched aerogel with $\delta_u=-0.03$. The polar-core vortex is stable at higher temperatures than the solid curve and no VCT occurs in the B-like phase. In the Inset, the total free energy for ${\tilde r}_b=30$ is compared between the nonunitary MHV (dashed curve) and the polar-core vortex (solid one) as a function of the temperature (mK). The VCT occurs at the temperature pointed by an arrow.\label{Fig:vt-abm}}
\end{figure}

\begin{figure}[t]
\includegraphics[scale=0.5]{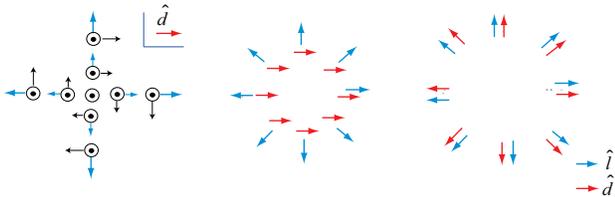}
\caption{(Color online) Sketch of the ${\bf l}$ texture (blue arrows) and the ${\bf d}$ texture (red ones) in the polar-core vortex far from the vortex center (right) and inside the bump in $f({\tilde r})$ (center). Inside the bump, order-parameter field can be expressed as $A_{\eta,i}\simeq \delta_{\eta,x}(a_{x,z}{\hat z}+i \, a_{x,\phi}{\hat\phi})_i$, indicating ${\bf l}\parallel {\hat r}$ and ${\bf d}\parallel{\hat x}$, where $a_{x,\phi}$ corresponds to $a_{x,y}$ for $\phi=0$ and $a_{x,x}$ for $\phi=\pi/2$ in Eq. (\ref{Eq:alike_xy}). The circle symbol and black arrow in the left figure expressing the close vicinity of the vortex center denote $a_{x,z}{\hat z}$ and $a_{x,\phi}{\hat \phi}$, respectively. In the left figure, on approaching the vortex center, $a_{x,\phi}$ (i.e., the length of the 
black arrow) gradually vanishes, and at the vortex center, the polar state $A_{\eta,i}=\delta_{\eta,x}(a_{x,z}{\hat z})_i$ is realized. \label{Fig:d-vector}}
\end{figure}

Finally, we comment on the origin of the bump seen in $f({\tilde r})$ of the polar-core vortex in Fig. \ref{Fig:edens}.   
As one can see in Fig. \ref{Fig:op-xy}(b), on approaching the vortex center along the $x$ axis [solid (black) curves], $A_{\eta ,i}(x)\sim \delta_{\eta,x}(i \, a_{x,y} \, \hat{y}+a_{x,z} \, \hat{z})_i$ is almost constant in ${\tilde r} \geq 4$, implying that ${\bf l}$ vector remains nearly parallel to the radial direction. In contrast, on approaching the core along the $y$ axis [dashed (red) curves], we have the crossover from $A_{\eta ,i}(y) \sim \delta_{\eta,y}(a_{y,x} \, \hat{x}+i \, a_{y,z} \, \hat{z})_i$ to $\delta_{\eta,x} (i \, a_{x,x} \, \hat{x}+a_{x,z} \, \hat{z})_i$ on sweeping through the bump in the free-energy density around ${\tilde r}=10$. The spacial variation in $A_{\eta,i}$ discussed above are schematically shown in Fig. \ref{Fig:d-vector}, where blue and red arrows denote ${\bf l}$ and ${\bf d}$, respectively. On approaching the vortex center (from right to center in Fig. \ref{Fig:d-vector}), the ${\bf d}$ vector suddenly rotates to have an alignment parallel to the $x$ axis while keeping ${\bf l}$ in the radial direction. This energy cost of the resulting ${\bf d}$ texture is the origin of the bump seen in Fig. \ref{Fig:edens}. Further, close to the vortex center, the orbital component perpendicular to the vortex axis gradually vanishes keeping ${\bf d}$ in the $x$ direction (left in Fig. \ref{Fig:d-vector}) and the polar pairing state $A_{\eta,i} = a_{x,z} \delta_{\eta,x} \delta_{i,z}$ occurs at the vortex center. Therefore, we can deduce that a possible polar-core vortex is accompanied by a nonaxisymmetric ${\bf d}$ texture which, in turn, removes a singularity due to the pure phase, while keeping the ${\bf l}$ vector oriented in the radial direction. This behavior of the order parameter, implying a hybrid texture of the ${\bf d}$ vector and the pure phase, resembles that of the half quantum vortex \cite{SV} which might be realized in the A-like phase in an uniaxially {\it compressed} aerogel \cite{Volovik}.

\section{summary and discussions}
Possible discontinuous VCTs have been studied for both A-like and B-like phases in superfluid $^3$He in aerogels with uniaxial deformation parallel to the vortex axis. One VCT should be realized in the B-like phase in a uniaxially compressed aerogel, and a different VCT between a nonunitary core and a polar core may occur in the A-like phase in a uniaxially stretched aerogel. The former transition should appear even in low-pressure limit while it should be absent in aerogels stretched along the vortex axis. This conclusion might be relevant to the fact that no VCT has been detected in a B-like phase in a previous rotating measurement \cite{Ishikawa}. By contrast, it may not be easy to experimentally detect the VCT in the stretched A-like phase because of the random orientation of ${\bf l}$ due to the quenched disorder effect of aerogel neglected here. Nevertheless, upon cooling the system through $T_c$ {\it while} rotating the aerogel including $^3$He, this VCT may be seen if the lattice consisting of radial MHVs (Ref. \cite{NFOT}) is formed in bulk $^3$He-A under the similar condition because the resulting vortex lattice in the close vicinity of $T_c$ consists of cores overlapping with one another so that the order-parameter configuration near the core dominates over the quenched disorder effect making the order-parameter random. An additional feature in vortices in anisotropic aerogels is the presence of a nonunitary core state which may be detected by performing a magnetization measurement.

\section{Acknowledgement}
The authors thank T. Takagi for useful discussions and O. Ishikawa for a comment. This work was supported by the Japan Society for the Promotion of Science.

\end{document}